\documentclass[conference]{IEEEtran}
\IEEEoverridecommandlockouts
\usepackage{cite}
\usepackage{amsmath,amssymb,amsfonts}
\usepackage{algorithmic}
\usepackage{graphicx}
\usepackage{textcomp}
\def\BibTeX{{\rm B\kern-.05em{\sc i\kern-.025em b}\kern-.08em
    T\kern-.1667em\lower.7ex\hbox{E}\kern-.125emX}}
\begin{document}

\title{IoT Based Real Time Noise Mapping System for Urban Sound Pollution Study}

\author{\IEEEauthorblockN{1\textsuperscript{st} Sakib Ahmed}
\IEEEauthorblockA{\textit{Department of Electrical and Electronic Engineering} \\
\textit{BRAC University}\\
Dhaka, Bangladesh \\
sakibbracueee@gmail.com}
\and
\IEEEauthorblockN{2\textsuperscript{nd} Touseef Saleh Bin Ahmed}
\IEEEauthorblockA{\textit{Department of Electrical and Electronic Engineering} \\
\textit{BRAC University}\\
Dhaka, Bangladesh \\
touseefsba@gmail.com}
\and
\IEEEauthorblockN{3\textsuperscript{rd} Sumaiya Jafreen}
\IEEEauthorblockA{\textit{Department of Electrical and Electronic Engineering} \\
\textit{BRAC University}\\
Dhaka, Bangladesh \\
sumaiyajafreenp@gmail.com}
\and
\IEEEauthorblockN{4\textsuperscript{th} Jannatul Tajrin}
\IEEEauthorblockA{\textit{Department of Computer Science and Engineering} \\
\textit{BRAC University}\\
Dhaka, Bangladesh \\
jannatultajrin33@gmail.com}
\and
\IEEEauthorblockN{5\textsuperscript{th} Jia Uddin}
\IEEEauthorblockA{\textit{Department of Computer Science and Engineering} \\
\textit{BRAC University}\\
Dhaka, Bangladesh \\
jia.uddin@bracu.ac.bd}
}

\maketitle

\begin{abstract}
This paper describes the development of a system that enables real time data visualization via a webapp regarding sound intensity using multiple node devices connected through internet. The prototypes were realized using ATmega328 (Arduino Nano) and ESP8266 hardware modules, NodeMCU Arduino wrapper library, Google maps and firebase API along with JavaScript webapp. System architecture is such that multiple node devices will be installed in different locations of the target area. On each node device, an Arduino Nano interfaced with a Sound Sensor measures ambient sound intensity and ESP8266 Wi-Fi module transmits the data to a database via web API. On the webapp, it plots all the real-time data from the devices over Google maps according to the locations of the node devices. The logged data that is collected can then be used to carry out researches regarding sound pollution in targeted areas.
\end{abstract}


\section{Introduction}

The world health organization (WHO) estimates that 800,000 people per year die from the effects of air pollution \cite{a1}. Disclosure to excessive noise levels is known to negatively impact quality of life. Sound pollution has effect on concentration, annoyance, communication disturbance, sleeping disruption, health issues and so on. While instances of sleep disruption and affected concentration (represented by interruption of an activity in response to a noise occurrence) can be measured, annoyance is determined based on the perception of given sounds or as a consequence of the former effects. Further to this, exposure to excessive noise levels is known to have detrimental health impacts (at sound pressure levels above 65 dBA) \cite{a2}. Noise pollution continues to pose a major health threat for urban areas, especially in cities and particularly in Dhaka city. In this city of 12 million people \cite{a3}, there is no concrete scalable system to measure sound intensity to deduce noise pollution. At the same time, Environmentalist do not have adequate data to carry out appropriate actions \cite{a4}.

Scenarios like this where mass data is needed from a spread locale of vast areas simultaneously, the IoT approach comes handy to reach a scalable solution \cite{a5}. As mentioned in \cite{a6}, IoT simplifies Human-Device interaction as well as Device-Device interaction. This particular development in the shape of new avenues of interactions will impact essentially every industry such as transportation and logistics, energy, healthcare etc. \cite{a6}.

Several research works have reputed in very similar aspects. An earlier approach to noise mapping proposed a smartphone based sensing \cite{a7}. In \cite{a8}, a network of sensors connected over ZigBee and Wi-Fi protocol was implemented for real-time sensor data affiliation where all the processing has been done in the online cloud server for factory environment monitoring. In \cite{a9}, proposed a very similar WSN model using the IoT architecture where they demonstrated using temperature and humidity sensors in the labs.

In the present paper, the proposed system for real time noise data mapping via IoT takes the best of each approach and combines with newer technologies. Incorporation of free and open source web services reduces the necessity of costly cloud services. The prototype of node devices consists ESP8266 module, due to the issues of simple implementation and also due to the fact, that the module is extremely affordable regarding the price in contrast to its features. It is therefore possible for the proposed system upgrade or extend on demand. The system should therefore provide the following capabilities:
\begin{enumerate}
\item Highly scalable due to lower cost.
\item Provide the scope for incorporating additional sensors
\end{enumerate}
The main motivation of this project is to create a system or platform so that these noise or environment data can be collected and sent to the right places where the information is needed. Thus proper countermeasures can be taken to locate and reduce noise pollution in areas where intervention is needed.
The reason for the proposed system to use IoT infrastructure is that there is no need of human interaction. They are interrelated computing devices having ability to transfer data over network without human-machine interaction.

The rest of this paper is organized as follows: Section 2 describes detail on the implementation of the system. The results and discussion is presented in section 3 while Section 4 concludes the paper.

\section{System Overview}

\begin{figure}[htbp]
\centerline{\includegraphics[width=0.45\textwidth]{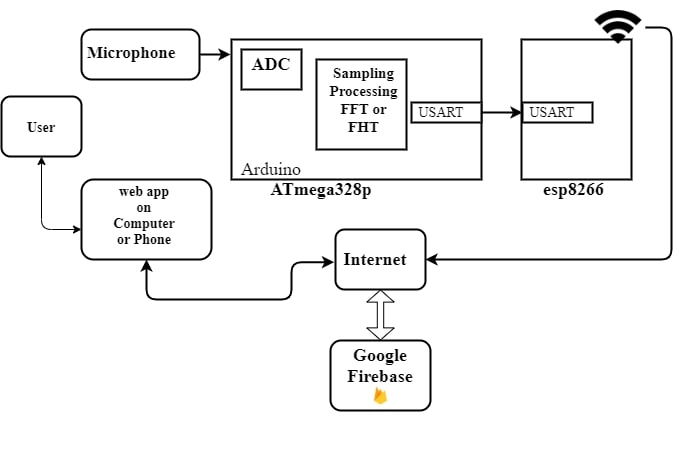}}
\caption{ Block Diagram.}
\label{fig:blockdiagram}
\end{figure}

The whole system can be segmented in hardware (Node devices), database and webapp. The block diagram in figure \ref{fig:blockdiagram} illustrates the system overview.

Analog reading from the microphone will be sent to the ADC of the ATmega328p microcontroller of the Arduino board. After the signal has been processed it is sent to the NodeMCU which with the help of Wi-Fi sends the processed data to the database which is Google Firebase via internet \cite{a9}. Firebase is basically a database which handles the real time data storing and serving. The real-time data is then pulled by the webapp and displayed on over google Maps in a web browser in terms of red to green gradient overlay. Red regions indicate high noise levels and green areas indicate quiet regions.

\subsection{Hardware}
\begin{figure}[htbp]
\centerline{\includegraphics[width=0.45\textwidth]{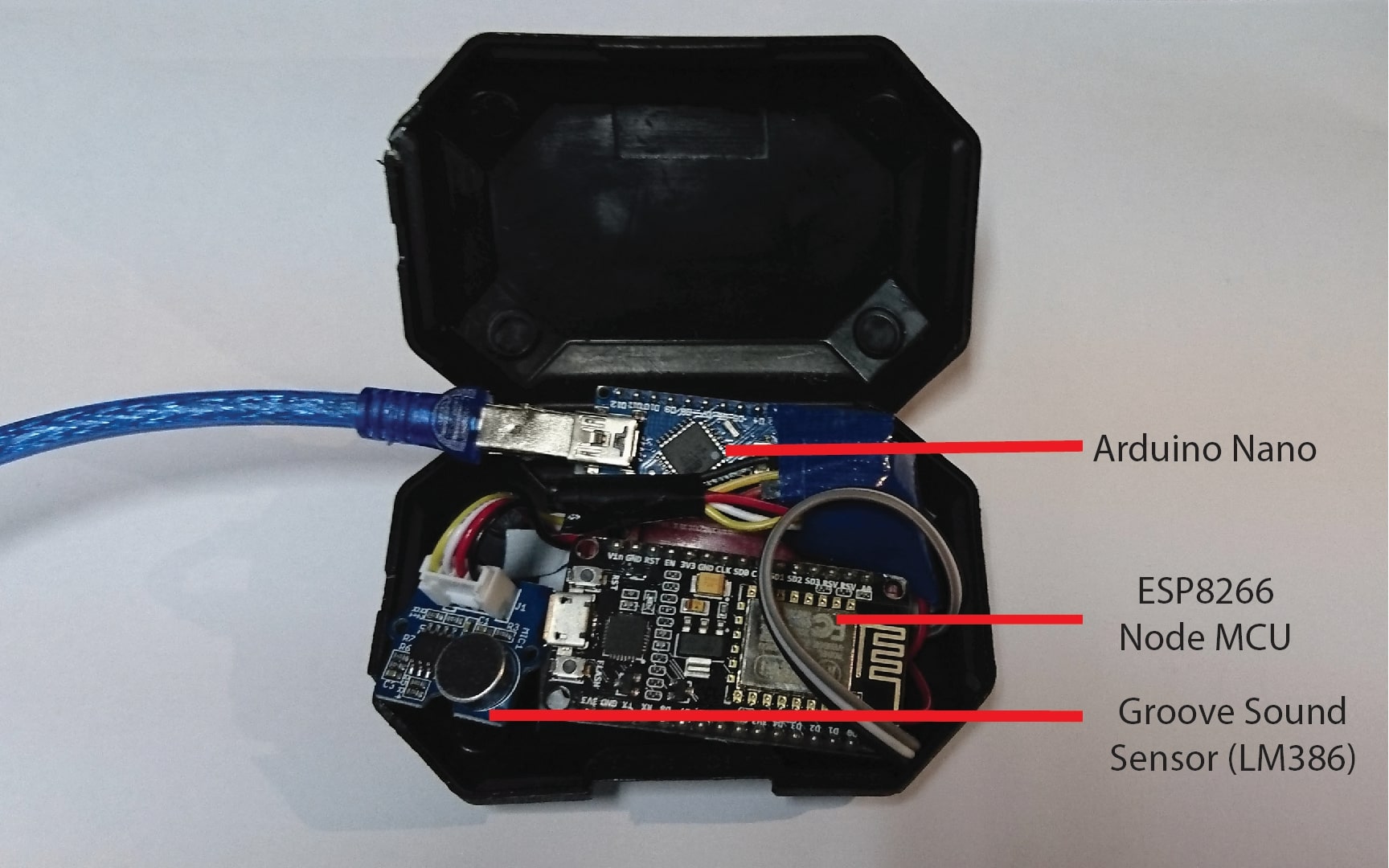}}
\caption{ Prototype}
\label{fig:prototype}
\end{figure}

Each node device consists of an Arduino Nano, a loudness/sound sensor and a NodeMCU (ESP8266) module, everything powered from a single 5V smartphone DC wall adopter or a smartphone power bank. In figure \ref{fig:prototype} is an image of one of the three node-device prototypes built for this project.

\subsubsection*{Sound Sensor} Sound Sensor detects the sound intensity of the environment and feeds to the ADC of Arduino. The main component of the module is a simple microphone, which is based on the LM386 amplifier and an electret microphone. This module is a 3pin sensor; power, ground and an output which is analog and can be easily interfaced and sampled by a microcontroller.

\subsubsection*{Arduino} 
\begin{figure}[htbp]
\centerline{\includegraphics[width=0.35\textwidth]{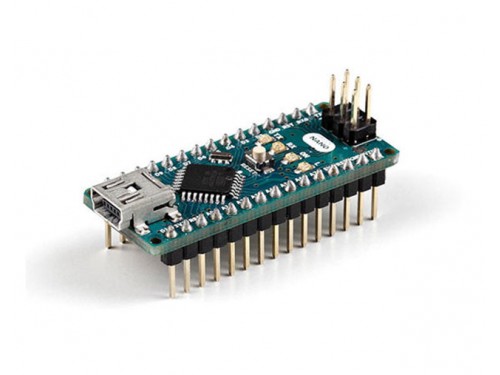}}
\caption{ Arduino Nano}
\label{fig:nano}
\end{figure}

Arduino Nano is the sound processing device in the proposed system. It’s based on ATmega328p chip which is an 8bit microcontroller. The 10bit ADC of Arduino is used for processing the analog sound data from the sensor where the input data from the microphone undergoes Sampling and processing by means of FFT (FHT) \cite{a7,a8}. After the signal has been processed it is sent via USART of the Arduino to the USART of the NodeMCU at a 2 second interval. Programming language used for this chip is Arduino Software (IDE), based on C++. Being relatively inexpensive the Arduino also simplifies the process of working with microcontrollers. Figure \ref{fig:nano} shows an Arduino Nano.

\subsubsection*{NodeMCU (ESP8266)} This is the heart of the node device. It provides the platform for IoT. It’s a Wi-Fi module with esp8266 firmware within. It takes processed sensor data from the Arduino and uploads to the database. The developer of this board is ESP8266 Open source Community. It runs on the NodeMCU operating system based on LUA scripting language. The CPU is ESP8266 (LX106). It has an in-built memory of 128 Kbytes and a storage capacity of 4 Mbytes. With a physical size of 49 x 24.5 x 13mm and an USB port connecting the computer, this chip provides around 0.00026W – 0.56W consumption of power. Attaining all the criteria this chip is so far the leading hardware around and is the future of IoT. In programming this device, Arduino wrapper/interpreter library is used which allowed programming using Arduino IDE in C++.
\begin{figure}[htbp]
\centerline{\includegraphics[width=0.35\textwidth]{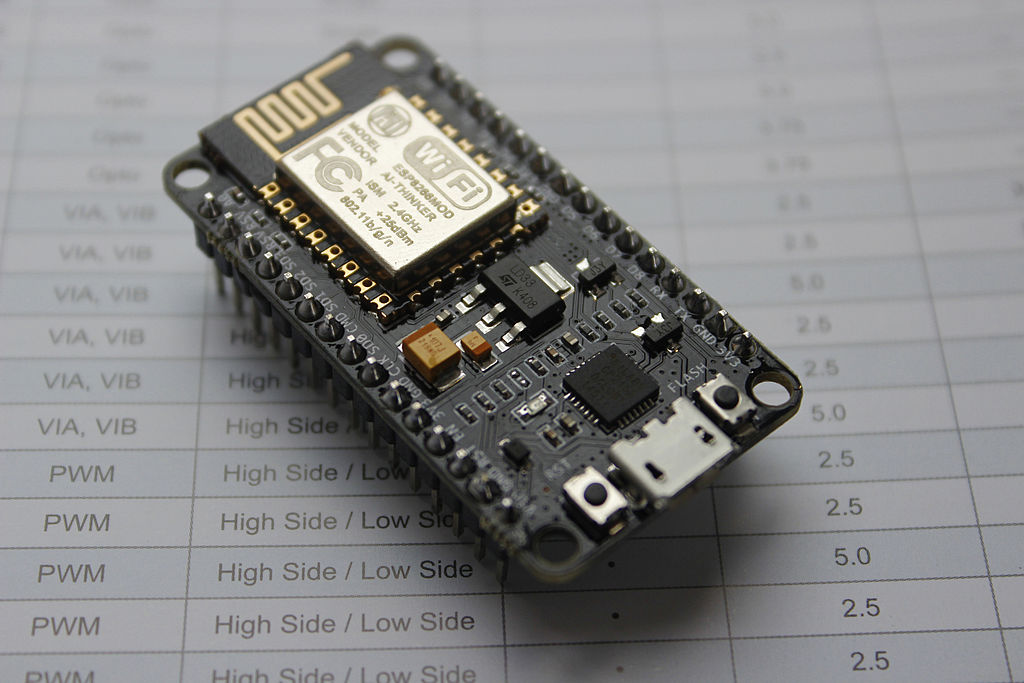}}
\caption{ ESP8266 (NodeMCU) hardware module }
\label{fig:esp}
\end{figure}

\subsection{Database}

For the data base of the proposed system, Google Firebase has been used. Firebase is a cloud storage service that we have leveraged in order to enable our middleware to be decoupled and interface with 3rd party applications \cite{a10}. Firebase was used since the model is similar to Google REST API. 

\begin{figure}[htbp]
\centerline{\includegraphics[width=0.35\textwidth]{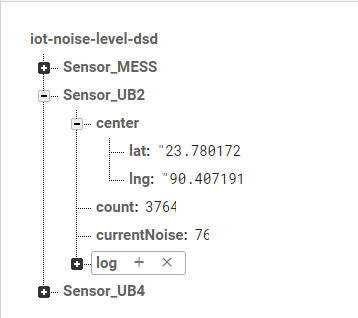}}
\caption{ Firebase web console, showing the data from 
the proposed system }
\label{fig:firebase}
\end{figure}

Firebase allows developers to create mobile and web applications that used the data generated by their smart thermostat and smoke detector, without having to adjust to the specific format of data they were generating. Figure \ref{fig:firebase} shows the web interface where data is stored on Firebase and can be edited. Since Firebase stores data in a JSON format, NodeMCU can already interpret that format \cite{a11}.

Firebase also allows developers to create listeners for specific sections of the JSON document that will fire when data is changed, added, removed or moved. This makes creating an asynchronous mobile or web application interface incredibly simple. Furthermore, since Firebase enables users to add authentication to their own personal Firebase, users can rest assured that their data is protected from malicious attackers.

\subsection{Webapp}

The webapp is developed in JavaScript which uses Google maps and Firebase JavaScript API. The real-time data is pulled by the webapp and displayed on over instance of Google Maps in a web browser in terms of red to green gradient overlay. Red regions indicate high noise levels and green areas indicate quiet regions.  

Every time current noise data is updated, the webapp asynchronously updates the graphical view in correspondence to the value.

\section{Result Analysis}
\begin{figure}[htbp]
\centerline{\includegraphics[width=0.9\linewidth]{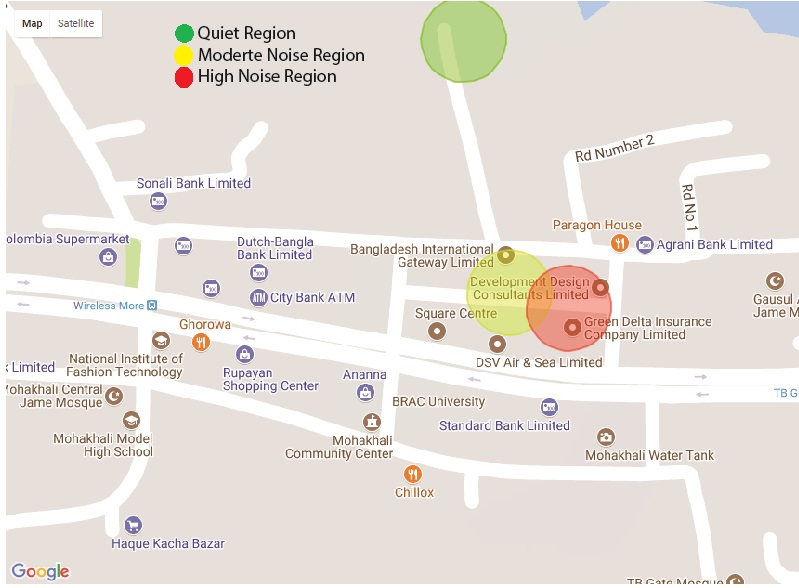}}
\caption{ Webapp’s real-time noise mapping }
\label{fig:map}
\end{figure}

The designed system incorporating three node devices was able to produce expected outcomes and display real-time noise mapping on the webapp. Figure \ref{fig:map} shows the webapp’s visual representation of the real-time noise mapping.

\begin{figure}[h]
\centerline{\includegraphics[width=0.80\linewidth]{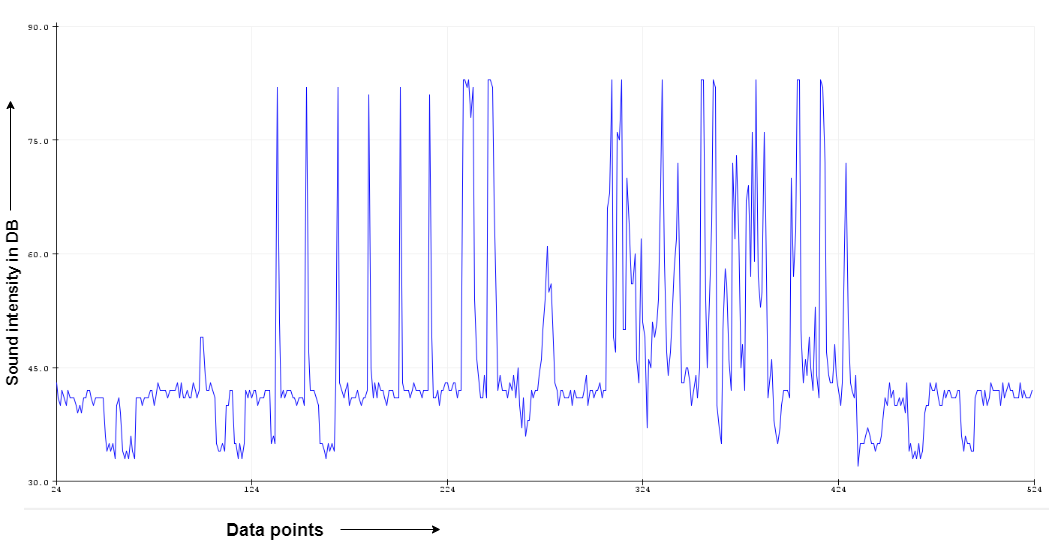}}
\caption{ Decibel Spectrum from device }
\label{fig:graph_serial}
\end{figure}

\begin{figure}[t]
\centerline{\includegraphics[height=0.5\linewidth]{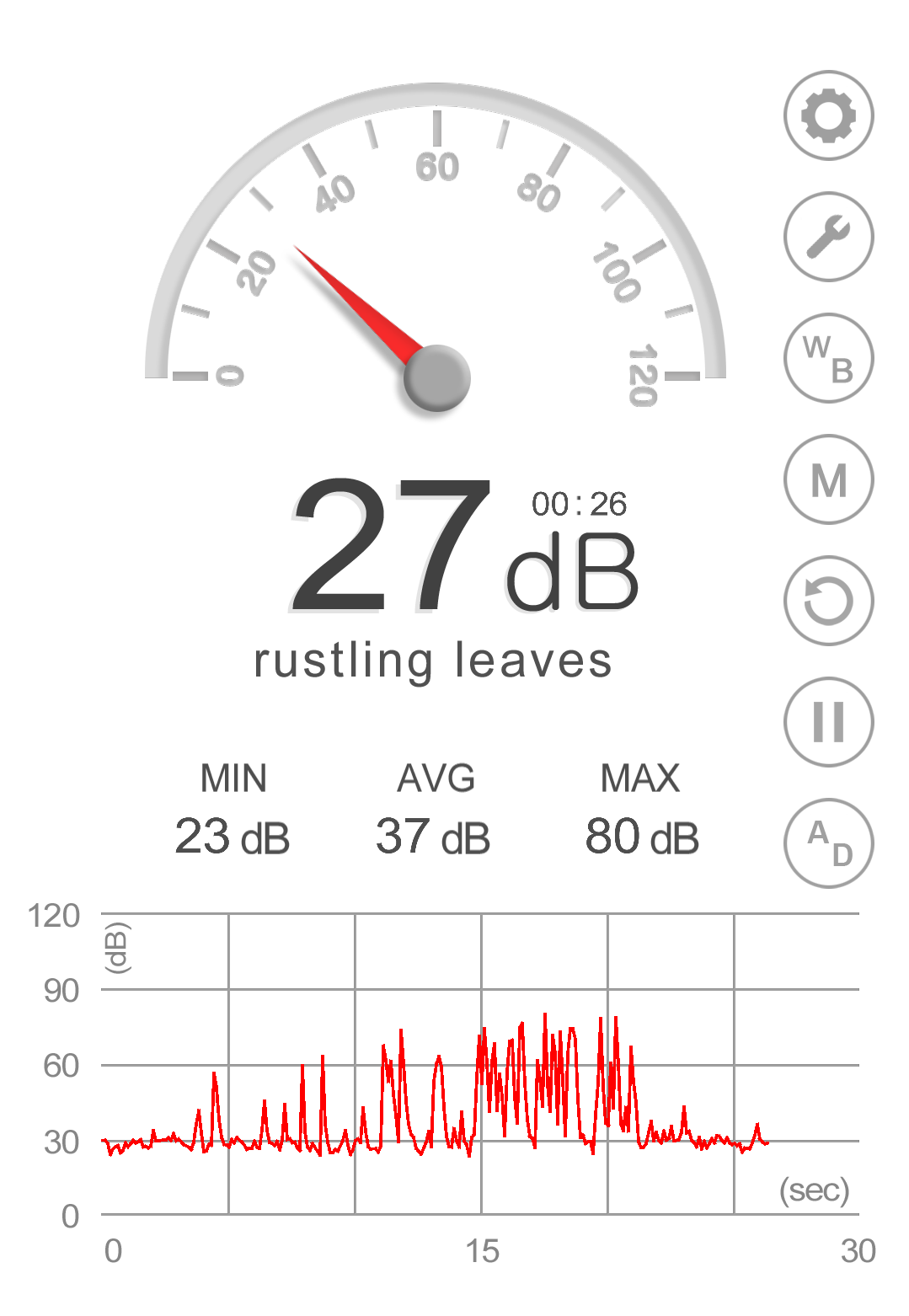}}
\caption{ Original Decibel Spectrum from a digital meter (Smart Phone App) }
\label{fig:graph_app}
\end{figure}

In the node devices, sound sampling data is shown in figure \ref{fig:graph_serial} taken from the Arduino IDE’s serial plotter and when compared with concurrent data from a digital decibel meter as shown in figure \ref{fig:graph_app}, shows significant similarities in the decibel graph indicating that the measurement being correct.
\section{Conclusion}

This device has both small scale as well as large scale applications. We can use it in industrial areas as there is a lot of noise pollution.  It can also be used in education systems to monitor the noise in classrooms. It can be used in heavily traffic regions as well as monitor noise in different parts of the city. In this paper one method of monitoring sound is presented using IoT. The device has the potential to significantly change the noise pollution monitoring system in Dhaka City. It will not require any human-machine interaction or human effort. This will create a research platform for environmentalists and other social activity groups and allow them to receive data easily and cheaply. The device is mainly using Arduino to process and store the data, sound sensor to take noise input and with the help of NodeMCU it is presenting the data in a website for people to observe. Implementation of this device will provide a low cost and well organized solution of monitoring environmental noise at real time.

\end{document}